\documentstyle[12pt,epsf]{article}

\begin{document}

\title{Exact effective action for N=1 supersymmetric theories.}

\author{P.~I.~Pronin
\thanks{E-mail: $petr@theor.phys.msu.su$}
and K.~V.~Stepanyantz
\thanks{E-mail: $stepan@theor.phys.msu.su$}}

\maketitle

\begin{center}
{\em Moscow State University, Physics Faculty,\\
Department of Theoretical Physics.\\
$117234$, Moscow, Russian Federation}
\end{center}

\begin{abstract}
We investigate nonperturbative effects in N=1 supersymmetric theories
and propose a new expression for the effective action, which correctly
reproduces quantum anomalies and agrees with the transformation law of
instanton measure. Actually the result is a nonperturbative extension of
Veneziano-Yankielowitch effective Lagrangian. The possibility of
integrating out the gluino condensate is discussed.
\end{abstract}


\section{Introduction}
\hspace{\parindent}

Investigation of nonperturbative dynamics is a very important problem
of quantum field theory. It is well known \cite{thooft,shifman}, that
except for perturbative corrections there is a series of instanton
contributions. Their sum was found exactly in \cite{seiberg} for N=2
supersymmetric Yang-Mills theory with $SU(2)$ gauge group in the constant
field limit. It was checked, that the asymptotic of exact result reproduced
one-instanton contribution correctly \cite{finnell,yung}.

Attempts to construct exact results for N=1 SUSY theories were made in
\cite{affleck,seiberg2}. However, the corresponding results do not
agree with instanton calculations. (In the case $N_f=N_c-1$ the agreement
takes place only at the one-instanton level \cite{finnell}, while higher
instanton corrections break it.) Moreover, Affleck-Dine-Seiberg results
do not produce correct expressions for anomalies, because they do not
contain any gauge degrees of freedom.

In principle, at the perturbative level the anomalies is correctly
reproduced by the Veneziano-Yankielowitch effective Lagrangian
\cite{veneziano}, containing gluino condensate. However, this result is
not applicable beyond the frames of perturbation theory.

The purpose of the present paper is to obtain the exact (nonperturbative)
effective Lagrangian for N=1 SUSY Yang-Mills theory with matter. It should
correctly reproduce anomalies, be in agreement with instanton calculations
(at least with the transformation law of the collective coordinate measure)
and have a structure similar to the N=2 case (because N=2 SUSY theories
have also N=1 SUSY).

In order to do it we will use the method, based on the consideration
of quantum anomalies \cite{anomalybook} beyond the frames of perturbation
theory. The first exact expression was found recently for R-anomaly in N=2
supersymmetric Yang-Mills theory \cite{matone,howe,sonnenschein,eguchi}.
Due to the instanton contributions it differs from the perturbative result.
\footnote{
Such possibility was pointed rather long ago \cite{shifmanomaly}, but a
series of instanton corrections with unknown coefficients produced
considerable difficulties.}

The derivation was based on the Seiberg and Witten exact expression
\cite{seiberg}, but the result appeared to have a very simple
interpretation: exact anomaly is a vacuum expectation value of the
perturbative one. Nevertheless, for checking this relation one should
essentially use the exact prepotential, found in \cite{seiberg} by
completely different methods. Thus, we come to the question, whether it is
possible to solve the inverse problem, i.e. to derive exact results from
the form of anomalies. This idea really allows to derive Seiberg-Witten
solution in N=2 SUSY $SU(2)$ Yang-Mills theory and the general structure
of Picard-Fuchs equations in other N=2 SUSY theories (first obtained in
\cite{add1,add2} by other methods) \cite{our}.

In the present paper the corresponding approach is applied to N=1
supersymmetric Yang-Mills theories.

Our paper is organized as follows:

In Section \ref{rel} we briefly discuss the relation between perturbative
and exact anomalies, which is used in Section \ref{section3} to derive the
exact effective Lagrangian. First, in Subsections \ref{general_structure}
we apply it (together with the results of Appendix \ref{instanton} and
\ref{derive_z}) to investigate the general structure of the superpotential.
The exact result is obtained in Subsection \ref{exact_result}. In Section
\ref{int_out} we discuss the possibility of integrating out the gluino
condensate and equivalence of the obtained expression and Seiberg's exact
results. Conclusion is devoted to the brief review and discussion of the
results. In the Appendix \ref{susy} we review the necessary information
concerning sypersymmetric theories and summarize our notations. Then, in
Appendix \ref{instanton} we investigate the structure of nonperturbative
corrections, that agrees with the transformation law of the collective
coordinate measure under $U(1)_x$ transformations. In Appendix \ref{moduli}
we briefly review the structure of moduli space for N=1 supersymmetric
theories, that is used in Appendix \ref{derive_z} to rewrite the effective
action in the gauge invariant form.


\section{The relation between perturbative and exact anomalies}
\label{rel}
\hspace{\parindent}

Now let us consider N=1 supersymmetric $SU(N_c)$ Yang-Mills theory
with $N_f$ matter supermultiplets and find the general structure of
the effective superpotential and anomalies. (Our notation are summarized
in Appendix \ref{susy}.)

The effective Lagrangian can be split into the following parts
\cite{veneziano}

\begin{equation}
L_{eff} = L_{k} +  L_{a} + L_m,
\end{equation}

\noindent
where

\begin{eqnarray}
&& L_k = \int d^4\theta\ K(S, S^{*},\Phi, \Phi^{*}); \nonumber\\
&& L_a = \mbox{Re} \int d^2\theta\ w(S,\Phi)
\end{eqnarray}

\noindent
and $S\equiv\mbox{tr} W^2$ is the gluino condensate.

Here $L_k$ denotes kinetic term, that does not contribute to the anomaly,
$L_{a}$ is a holomorphic part of the superpotential and $L_m$ is a mass
term. Below we will consider only massless case ($L_m=0$). Therefore,
the only nontrivial contribution to anomalies comes from $L_a$ and it is
the only part, that we are able to investigate. (Our method can not give
any information about a possible kinetic term.)

In order to define $L_a$ we consider the anomaly of $U(1)_x$-symmetry
(for more details see Appendix \ref{susy})

\begin{eqnarray}
U(1)_x: &&W(\theta) \to e^{i\alpha} W(e^{-i\alpha\gamma_5}\theta);\nonumber\\
&&\phi(\theta)\to e^{ix\alpha} \phi(e^{-i\alpha\gamma_5}\theta);\nonumber\\
&&\tilde\phi(\theta)\to e^{ix\alpha} \tilde\phi(e^{-i\alpha\gamma_5}\theta)
\end{eqnarray}

\noindent
beyond the frames of perturbation theory. The result can be found by
using the relation between perturbative and exact anomalies.

Of course, exact anomalies are quite different from perturbative ones.
For example, the exact expression for R-anomaly in the N=2 SUSY $SU(2)$
Yang-Mills theory, found by transforming Seiberg-Witten effective action
\cite{matone,bellisai}, is

\begin{equation}\label{npa}
\langle\partial_\mu j_R^\mu\rangle =
\frac{1}{16\pi} \mbox{Re} \int d^2\theta_1 d^2\theta_2
\Big(F+F_D \Big) =
\frac{1}{8\pi^2} \mbox{Im} \int d^2\theta_1 d^2\theta_2 u.
\end{equation}

\noindent
where $u\equiv {\displaystyle \frac{1}{2}}\langle\Phi^2\rangle$ and $\Phi$
is N=2 superfield

\begin{eqnarray}
&&\Phi(y,\theta_1,\theta_2) = \phi(y,\theta_1) - i \bar\theta_2(1+\gamma_5)
W(y,\theta_1) + \frac{1}{2} \bar\theta_2 (1+\gamma_5) \theta_2 G(y,\theta_1);
\nonumber\\
&&y^\mu = x^\mu
+{\displaystyle \frac{i}{2}} \bar\theta_i \gamma^\mu\gamma_5\theta_i;
\nonumber\\
&&G(y,\theta_1) = \frac{1}{2} \int d^2\bar \theta_1 e^{2V} \phi^{+} e^{-2V}.
\end{eqnarray}

Here we should attract attention to the easily verified identity

\begin{equation}\label{identity}
F+F_D= - \frac{2i}{\pi} u
\end{equation}

\noindent
that will be used below.

From the other side, in the perturbation theory

\begin{equation}\label{pa}
\mbox{\bf A}\equiv \langle\partial_\mu j_R^\mu\rangle_{pert} =
\frac{1}{16\pi^2} \mbox{Im}\ \mbox{tr}\int d^2\theta_1 d^2\theta_2 \Phi^2.
\end{equation}

Of course, the expressions (\ref{npa}) and (\ref{pa}) are quite different.
The former is a series over $\Lambda^4$ produced by instanton contributions.
In particular, taking into account one instanton correction we have
\cite{finnell,yung}

\begin{equation}\label{correct}
\langle\partial_\mu j_R^\mu\rangle
= \frac{1}{16\pi^2} \mbox{Im}\ \mbox{tr}\int d^2\theta_1 d^2\theta_2
\left[\Phi^2 +\frac{\Lambda^4}{2\Phi^2} +O(\Lambda^8)\right],
\end{equation}

\noindent
that in components can be written as

\begin{eqnarray}
&&\langle\partial_\mu j_R^\mu\rangle
=\frac{e^2}{4\pi^2} \Big(1 + \frac{3\Lambda^4}{2 e^4 \varphi^4}\Big)
F_{\mu\nu} \tilde F^{\mu\nu}
-\frac{3\Lambda^4}{e^2 \pi^2 \varphi^5}F_{\mu\nu}
\bar\Psi_D \Sigma_{\mu\nu}\gamma_5 \Psi_D\nonumber\\
&&\qquad\qquad\qquad\qquad\qquad\qquad\qquad\qquad
+\frac{60\Lambda^4}{e^2 \pi^2 \varphi^6}
(\bar\Psi_D\Psi_D) (\bar\Psi_D\gamma_5 \Psi_D)+O(\Lambda^8),\qquad
\end{eqnarray}

\noindent
where

\begin{equation}
\tilde F^{\mu\nu} = \frac{1}{2} \varepsilon^{\mu\nu\alpha\beta}
F_{\alpha\beta}
\end{equation}

\noindent
and we introduced a Dirac spinor

\begin{equation}
\Psi_D = \frac{1}{2}(1+\gamma_5)\psi_1+\frac{1}{2}(1-\gamma_5)\psi_2.
\end{equation}

And nevertheless, the nonperturbative result is only a vacuum expectation
value of the perturbative one, that in particular produces a natural
solution of anomalies cancellation problem in the realistic models.

This result is not unexpected. Really, performing, for example, chiral
transformation in the generating functional we have

\begin{eqnarray}
&&0 = \left.\frac{1}{Z} \frac{\delta Z}{\delta\alpha}\right|_{\alpha=0}
=\left.\frac{1}{Z} \frac{\delta}{\delta\alpha} \int DA D\bar\psi' D\psi'
\mbox{exp} \Big(i S - \partial_\mu\alpha j^\mu_5 \Big)\right|_{\alpha=0}
\nonumber\\
&&=\left.\frac{1}{Z} \frac{\delta}{\delta\alpha} \int DA D\bar\psi D\psi
\mbox{exp} \Big(i S - \partial_\mu\alpha j^\mu_5 - \alpha\mbox{\bf A}\Big)
\right|_{\alpha=0}
=\langle  \partial_\mu j^\mu_5 - \mbox{\bf A}  \rangle,
\end{eqnarray}

\noindent
where {\bf A} denotes the perturbative anomaly, produced by the measure
noninvariance \cite{fujikava}. Finally

\begin{equation}\label{relation}
\langle \partial_\mu j^\mu_5\rangle = \langle\mbox{\bf A}  \rangle.
\end{equation}

\noindent
(R-transformation are considered similarly).

It is just the relation, mentioned above. Of course, it is valid for a wide
range of models and is really a point to start with. Let us note, that the
derivation presented in \cite{matone} essentially used the form of
exact results. So, we are tempted to reverse the arguments. In the next
section we will try to apply this approach to N=1 SUSY theories.


\section{Exact effective Lagrangian}
\label{section3}

\subsection{General structure}\label{general_structure}
\hspace{\parindent}

Let us apply (\ref{relation}) to N=1 supersymmetric $SU(N_c)$ Yang-Mills
theory with $N_f$ matter supermultiplets. At the perturbative level
the anomaly of $U(1)_x$-symmetry has the following form

\begin{eqnarray}\label{pertanomaly}
&&\partial_\mu J^\mu_x = \Big(-N_f + N_c + x N_f\Big) \frac{1}{16\pi^2}
\varepsilon^{\mu\nu\alpha\beta} \mbox{tr} F_{\mu\nu} F_{\alpha\beta}
\nonumber\\
&&\qquad\qquad\qquad\qquad\qquad\qquad
= - \Big(N_f - N_c - x N_f\Big) \frac{1}{16\pi^2}
\mbox{Im\ tr}\int d^2\theta\ W^2,
\end{eqnarray}

\noindent
so that the exact anomaly is

\begin{eqnarray}\label{N1anomaly}
\langle \partial_\mu J^\mu_x\rangle
= - \Big(N_f - N_c - x N_f\Big) \frac{1}{16\pi^2}
\mbox{Im}\int d^2\theta\ u
\end{eqnarray}

\noindent
where $u\equiv \langle \mbox{tr} W^2\rangle$. Therefore, the effective
Lagrangian should depend in particular on $S= \mbox{tr} W^2$. This
result is not new. At the perturbative level the similar investigation
was made in \cite{veneziano}. However, in this paper we do not intend to
restrict ourselves by the frames of perturbation theory. Therefore, we
can not assume, that $u = \mbox{tr} W^2$ (Here we would like to remind
(\ref{correct})).

Taking into account the dependence on the gluino condensate and performing
$U(1)_x$ transformation in the effective action, from the other side we
obtain

\begin{equation}\label{N1anomaly5}
\langle\partial_\mu J^\mu_x\rangle =-\frac{\partial\Gamma}{\partial\alpha}
=- \mbox{Im} \int d^2\theta \Big(2w-2\frac{\partial w}{\partial S} S
- x \frac{\partial w}{\partial v} v\Big).
\end{equation}

\noindent
Comparing (\ref{N1anomaly}) with (\ref{N1anomaly5}) and taking into account,
that the equality should be satisfied for all $x$, we obtain

\begin{eqnarray}\label{2anomaly}
&&2w-2\frac{\partial w}{\partial S}S = \frac{1}{16\pi^2}(N_f-N_c) u;
\nonumber\\
&&\frac{\partial w}{\partial v} v = \frac{1}{16\pi^2} N_f u.
\end{eqnarray}

\noindent
This equation is very similar to the results of \cite{veneziano}.
Nevertheless, there is a crucial difference: $u\ne S$. Therefore,
one can only conclude that

\begin{equation}\label{ranom}
2w -2 \frac{\partial w}{\partial S} S
- \frac{N_f-N_c}{N_f}\frac{\partial w}{\partial v} = 0.
\end{equation}

This equation corresponds to the exact conservation of R-symmetry
at nonperturbative level. The similar condition was used in
\cite{affleck,seiberg2}, although the dependence $w=w(S)$ was ignored.
Of course, it is quite clear, that integrating $S$ out yields ADS
superpotential \cite{intriligator} and corresponds to imposing the
condition

\begin{equation}\label{condit}
\frac{\partial w}{\partial S} = 0.
\end{equation}

\noindent
Nevetheless, this equation can have no solutions. In this case the gluino
condensate can not be integrated out of the effective action and it is
impossible to obtain Seiberg's exact results. Below we will discuss
equation (\ref{condit}) in details.

It is desirable, that the solution of (\ref{ranom}) agrees with instanton
calculations. The necessary condition of it is the agreement with the
transformation law of the instanton measure. The presence of the gluino
condensate in the effective action allows to achieve it. The possible
structure of instanton corrections is analysed in the Appendix
\ref{instanton}. The result for n-instanton correction has the following
form

\begin{equation}\label{superpotential1}
w^{(n)}
=  S g_n\left(\frac{v^3}{S}\right)
\left(\frac{\Lambda}{v}\right)^{n(3N_c-N_f)}
\end{equation}

\noindent
where $g_n$ is an arbitrary function.

It can be easily verified, that the only solution of (\ref{ranom}),
agreeing with (\ref{superpotential1})\footnote{It corresponds to
${\displaystyle g_n(x) = \frac{1}{32\pi^2} c_n x^{n(N_c-N_f)},
\quad n\ge 1}$
in (\ref{superpotential1}).}, is

\begin{equation}\label{solution}
w = - \frac{1}{32\pi^2} S f(z);\qquad
f(z) = f_{pert}(z) + \sum\limits_{n=1}^\infty c_n z^n,
\end{equation}

\noindent
where

\begin{equation}
z\equiv\frac{\Lambda^{3N_c-N_f}}{v^{2N_f} S^{N_c-N_f}}
\end{equation}

\noindent
is a dimensionless parameter.
\footnote{Therefore, for $N_c=N_f+1$ first instanton correction does not
contain gluino condensate, that allows to compare it with ADS-superpotential
\cite{finnell}. However, higher corrections depend on $S$ and destroy the
agreement. So, the statement, that in this case instantons generate
ADS-superpotential is not correct.}

In the final result $z$ should be written in terms of gauge invariant
variables. Of course, the result will depend on the structure of moduli
space, that is briefly reviewed in Appendix \ref{moduli}. The derivation,
made in Appendix \ref{derive_z}, gives

\begin{eqnarray}\label{zz}
&&\hspace{-7mm}
z=\frac{\Lambda^{3N_c-N_f}}{\mbox{det} M\ S^{N_c-N_f}}, \qquad N_f<N_c;
\nonumber\\
&&\hspace{-7mm}
z=\frac{\Lambda^{3N_c-N_f}S^{N_f-N_c}}{
\mbox{det} M - (\tilde B^{A_1 A_2\ldots A_{N_f-N_c}} M_{A_1}{}^{B_1}
M_{A_2}{}^{B_2} \ldots M_{A_{N_f-N_c}}{}^{B_{N_f-N_c}}
B_{B_1 B_2\ldots B_{N_f-N_c}})},\nonumber\\
&&\hspace{110mm}
N_f\ge N_c.
\end{eqnarray}

In order to define $f_{pert}$ we note, that at the perturbative level $u=S$.
Therefore, in this case (\ref{2anomaly}) gives

\begin{equation}\label{fpert}
\frac{\partial f_{pert}}{\partial z} z = 1,
\end{equation}

\noindent
so that

\begin{equation}\label{expansion}
w = - \frac{1}{32\pi^2} S \Big(\ln z +
\sum\limits_{n=1}^\infty c_n z^n\Big).
\end{equation}

\noindent
Substituting it to (\ref{2anomaly}), we obtain

\begin{equation}\label{u_expansion}
u = S\Big(1 + \sum\limits_{n=1}^\infty n c_n z^n \Big),
\end{equation}

\noindent
that defines all anomalies in the theory according to (\ref{N1anomaly}).
At the perturbative level both (\ref{expansion}) and (\ref{u_expansion})
are certainly in agreement with \cite{veneziano}.


\subsection{Exact result}\label{exact_result}
\hspace{\parindent}

Let us define $f$ exactly. The general structure of the holomorphic
superpotential, found in section \ref{general_structure}, is similar to the
structure of the nonperturbative prepotential in the N=2 supersymmetric
Yang-Mills theory \cite{seiberg1}. In the latter case the relation between
perturbative and nonperturbative anomalies leads to Picard-Fuchs equations
\cite{our}, that can be used for derivation of exact results. Is it
possible to extend this approach to the case of N=1 supersymmetry?

First we substitute (\ref{solution}) into (\ref{2anomaly}), that gives

\begin{equation}\label{forf}
S \frac{df}{dz} z = u
\end{equation}

\noindent
(and therefore $u/S$ depends only on $z$).

The way to solve this equation is indicated by the analogy with N=2
supersymmetric $SU(2)$ Yang-Mills theory. In terms of N=1 superfields its
action is written as

\begin{equation}\label{N2versusN1}
\frac{1}{16\pi} \mbox{Im} \int d^4x d^2\theta \left(
\frac{d^2 F}{d\phi^2}W^2
+ \frac{1}{2}\int d^2\bar\theta \frac{dF}{d\phi}\phi^{+}\right).
\end{equation}

Let us compare it with

\begin{equation}\label{Sa}
S_a = - \frac{1}{32\pi^2} \mbox{Re}\int d^4x d^2\theta\ S f(z)
\end{equation}

\noindent
and introduce $a\equiv z^{- 1/4}$ (this choice of the power will be
explained below). The first term in (\ref{N2versusN1}) will coincide
with (\ref{Sa}) if

\begin{equation}\label{new}
2\pi i \frac{d^2F}{da^2}\equiv f;
\qquad \frac{d^2U}{da^2}\equiv \frac{u}{S}.
\end{equation}

\noindent
Then (\ref{forf}) takes the form

\begin{equation}\label{iden}
F+F_D= - \frac{2i}{\pi} U,
\end{equation}

\noindent
where

\begin{equation}
F_D=F-a a_D;\qquad a_D=\frac{dF}{da}.
\end{equation}

\noindent
This equation coincides with (\ref{identity}) and, therefore, we are
tempted to identify $F$ with Seiberg-Witten solution. It is so. Really,
differentiating (\ref{iden}) with respect to $U$, we obtain

\begin{equation}
a_D \frac{da}{du} - a \frac{da_D}{du} = - \frac{2i}{\pi}.
\end{equation}

\noindent
It means, that $a$ and $a_D$ are 2 independent solutions of the Picard-Fuchs
equation

\begin{equation}\label{pf}
\left(\frac{d^2}{da^2}+ L(U)\right)\left(\begin{array}{c}a\\a_D\end{array}
\right)=0,
\end{equation}

\noindent
where $L(U)$ is an undefined function.

At the perturbative level (see (\ref{fpert}))

\begin{equation}
\begin{array}{ll}
{\displaystyle f_{pert} = -4 \ln a\vphantom{\frac{1}{2}};}\\
{\displaystyle u_{pert} = W^2=S\vphantom{\frac{1}{2}},}
\end{array}
\qquad \mbox{so that}\qquad
\begin{array}{ll}
{\displaystyle F = \frac{i}{\pi} a^2
\Big(\frac{3}{2} - \ln a\Big);}\\
{\displaystyle U = a^2/2\vphantom{\frac{1}{2}}}
\end{array}
\end{equation}

\noindent
and, therefore

\begin{eqnarray}
&& a=\sqrt{2 U};\nonumber\\
&& a_D = - \frac{2i}{\pi} (a\ln a - a) =
- \frac{i}{\pi} \sqrt{2 U} \Big(\ln (2U) - 2\Big)
\end{eqnarray}

\noindent
satisfy

\begin{equation}
\left(\frac{d^2}{dU^2}+\frac{1}{4 U^2} \right)
\left(\begin{array}{c}a\\a_D\end{array}\right) = 0.
\end{equation}

However, the perturbative solution does not satisfy the requirement
\cite{seiberg,bilal}

\begin{equation}
\mbox{Im}\ \tau > 0,\qquad \mbox{where} \qquad
\tau = \frac{d^2F}{da^2} = \frac{da_D}{da} =\frac{1}{2\pi i} f,
\end{equation}

\noindent
that is derived exactly as in the N=2 case. Therefore, two singularities
(at $U=0$ and $U=\infty$) are impossible.

To find the structure of singularities let us note, that the solution
(\ref{expansion}) should contain all positive powers of z and, therefore,
is invariant under $Z_4$ transformations $a \to e^{i\pi k/2} a$. Taking into
account (\ref{new}) and (\ref{u_expansion}) we conclude, that the
corresponding transformations in the $U$-plane are $U \to e^{i\pi k} U$.
Thus, singularities of $L(U)$ in the Picard-Fuchs equation (\ref{pf}) should
come in pairs: for each singularity at $U=U_0$ there is another one at
$U=-U_0$.

Therefore, the considered model is completely equivalent to N=2
supersymmetric SU(2) Yang-Mills theory without matter and the only
possible form of Picard-Fuchs equation (up to the redefinition of
$\Lambda$) is

\begin{equation}\label{N1pf}
\left(\frac{d^2}{dU^2}+\frac{1}{4 (U^2-1)} \right)
\left(\begin{array}{c}a\\a_D\end{array}\right) = 0
\end{equation}

\noindent
with the solution \cite{seiberg,bilal}

\begin{equation}\label{final}
a(U)=\frac{\sqrt{2}}{\pi}\int\limits_{-1}^{1} dx
\frac{\sqrt{x-U}}{\sqrt{x^2-1}};
\qquad
a_D(U)=\frac{\sqrt{2}}{\pi}\int\limits_{1}^{U} dx
\frac{\sqrt{x-U}}{\sqrt{x^2-1}}.
\end{equation}

\noindent
Its uniqueness and, therefore, the uniqueness of the choice (\ref{N1pf})
was proven in \cite{unique}.

The function $F$ can be found by

\begin{equation}
\frac{dF}{dU} = a_D \frac{da}{dU}.
\end{equation}

\noindent
Its general structure is well known to be

\begin{equation}
F = - \frac{i}{\pi} a^2\Big(\ln a
+ \sum\limits_{n=0}^\infty F_n a^{-4n}\Big),
\end{equation}

\noindent
so that

\begin{equation}
f = 2\pi i \frac{d^2 F}{da^2} =
-4\ln a + \sum\limits_{n=0}^\infty f_n a^{-4n} =  \ln z +
\sum\limits_{n=0}^\infty f_n z^n.
\end{equation}

And now it is quite clear, that the choice $a = z^{-1/4}$ was made to
obtain the true structure of instanton corrections (\ref{solution}).

So, our main result is

\begin{equation}\label{result}
L_a = \frac{1}{16\pi} \mbox{Im} \int d^2\theta\
S \tau(z^{-1/4})
\end{equation}

\noindent
where $\tau(a)$ is Seiberg-Witten solution and $z$ is given by (\ref{zz}).


\section{On the impossibility to integrate out the gluino condensate
for $N_c>N_f$}\label{int_out}
\hspace{\parindent}

In this Section we will discuss the possibility of integrating the
gluino condensate out of the exact superpotential. Substituting
(\ref{result}) into the condition

\begin{equation}
\frac{\partial w}{\partial S} = 0
\end{equation}

\noindent
we can rewrite the latter in the following form ($a=z^{-1/4}$)

\begin{equation}\label{equ}
\frac{d\ln a}{d\ln\tau} = \frac{1}{4}(N_f-N_c)
\end{equation}

\noindent
Taking the perturbative asymptotic of the exact result we find, that

\begin{equation}
\ln(2a) = \frac{1}{4} (N_f-N_c)
\end{equation}

Of course, this equation has solution for all values of $N_f$ and $N_c$.
Therefore, at the perturbative level the gluino condensate can be always
integrate out, as it is usually assumed \cite{intriligator}. However, the
situation is quite different beyond the frames of the perturbation theory.
For the investigation it is very convenient to rewrite Seiberg-Witten
solution in terms of elliptic functions \cite{alvares}

\begin{equation}
a(u) = \frac{4}{\pi k} E(k);\qquad
a_D(u) = \frac{4}{i\pi k} \Big(E'(k)-K'(k)\Big)
\end{equation}

\noindent
where $k^2 = 2/(1+u)$. The functions $E$, $K$, $E'$ and $K'$ are defined
as

\begin{eqnarray}
&& E(k)
=\frac{\pi}{2}F(-\frac{1}{2},\frac{1}{2},1,k^2)
=\int\limits_0^1 dt \frac{\sqrt{1-k^2 t^2}}{\sqrt{1-t^2}}
\nonumber\\
&& K(k)
=\frac{\pi}{2}F(\frac{1}{2},\frac{1}{2},1,k^2)
=\int\limits_0^1 dt \frac{1}{\sqrt{(1-k^2 t^2)(1-t^2)}}
\nonumber\\
&& E'(k)=E(\sqrt{1-k^2});\qquad K'(k)=K(\sqrt{1-k^2})
\vphantom{\int\limits_0^1 }
\end{eqnarray}

\noindent
where $F$ is a hypergeometric function. The function $\tau$ is then given
by

\begin{equation}
\tau = \frac{i K'}{K}
\end{equation}

This form is very convenient for the computer research. Performing this
work we used the analytical calculation system "MAPPLE".

\begin{figure}[h]
\begin{picture}(0,0)(0,0)
\put(6.9,12.5){1.}
\put(4.5,12.5){2.}
\end{picture}
\epsfxsize15.0truecm\epsfbox{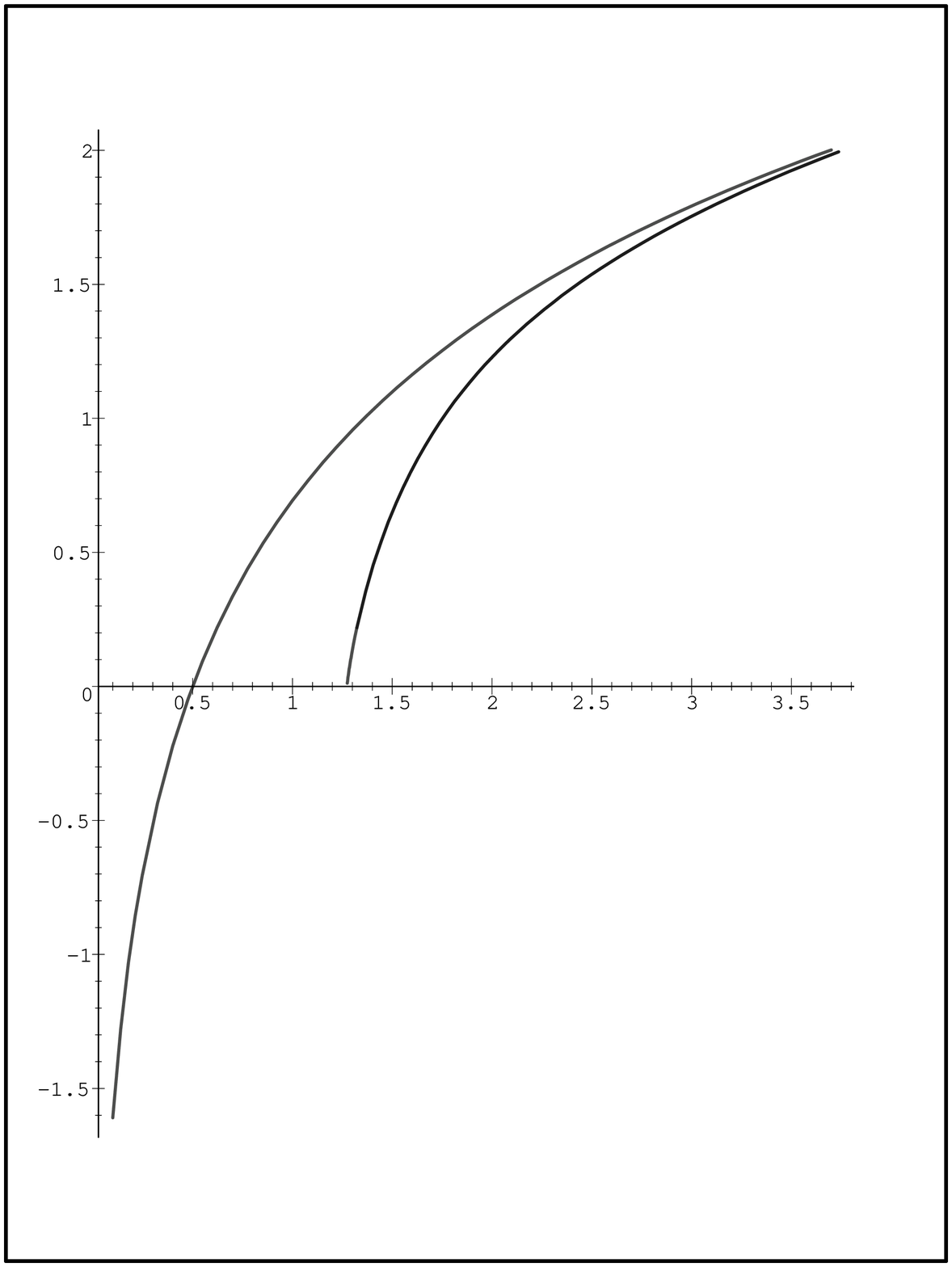}
\caption{Plots of the left hand side of the equation
${\displaystyle \frac{d\ln a}{d\ln\tau}=\frac{1}{4}(N_f-N_c)}$ as a
functions of the variable $a$. Curve 1. corresponds to the exact
result and curve 2. - to the perturbative one.}
\label{fig}
\end{figure}

The function

\begin{equation}\label{function}
\frac{d\ln a}{d\ln\tau}
\end{equation}

\noindent
is plotted at the Fig.\ref{fig} (curve 1.). For comparisons at this figure
we also present its perturbative asymptotic (curve 2.).

For $a\to\infty$ the perturbative result almost coincides with the exact
one. However, there are crucial differences for small $a$. Elliptic
functions are real only for $k^2=2/(1+u)<1$. In a region, where the
perturbation theory is not applicable ($k^2\to 0$), (\ref{function}) is
also positive. It is equal to 0 for $k^2=0$, which corresponds $a=4/\pi$.
Therefore, the range of values of (\ref{function}) is $[0,\infty)$ and
equation (\ref{equ}) has a solution only for a positive RHS, i.e. $N_c \le
N_f$.

For $N_c=N_f$ the classical constrain $\mbox{det} M - \tilde B B = 0$
is broken by instanton corrections \cite{seiberg2} as

\begin{equation}
\mbox{det} M - \tilde B B = \mbox{const}\ \Lambda^{2 N_f}
\end{equation}

In the frames of our approach it is produced automatically, because
in this case

\begin{equation}
L_a = \frac{1}{16\pi} \mbox{Im} \int d^2\theta\
S \tau(\Big(\frac{\Lambda^{2 N_f}}{\mbox{det} M - \tilde B B}\Big)^{-1/4})
\end{equation}

\noindent
and, therefore, the gluino condensate is a natural Lagrange multiplier.
Integrating it out, we obtain the equation $\tau(a) = 0$ which has a
solution

\begin{equation}
a = \Big(\frac{\Lambda^{2 N_f}}{\mbox{det} M - \tilde B B}\Big)^{-1/4}
= \frac{4}{\pi}
\end{equation}

\noindent
Finally, we obtain, that in this case

\begin{equation}
\mbox{det} M - \tilde B B = \Big(\frac{4}{\pi}\Big)^4\Lambda^{2 N_f}
\end{equation}


\section{Conclusion.}
\hspace{\parindent}

In the present paper we obtain the exact (nonperturbative) effective
Lagrangian (\ref{result}) for N=1 SUSY Yang-Mills theories with matter.
Our result has the following differences from the ones, found in
\cite{affleck,seiberg2}: Firstly, it agrees with the transformation law
of the collective coordinate measure under the chiral symmetries, due to
the presence of gluino condensate $S$. Secondly, the result correctly
reproduces anomalies beyond the frames of perturbation theory. And, finally,
it has the same structure as the Seiberg-Witten solution.

To derive the exact expression we developed a method, based on the relation
between perturbative and exact anomalies. Here we should mention, that the
similar approach was presented very long ago by Veneziano and Yankielowitch
\cite{veneziano}. Nevertheless, their derivation is valid only in the
perturbation theory. We would like to attract the attention, that
Veneziano-Yankielowitch effective Lagrangian is not applicable at the
nonperturbative level.

So, our result can be considered as a syntheses of Veneziano-Yankielowitch
effective Lagrangian and Seiberg's exact results, which is free from some
of their shortcomings.

\vspace{1cm}

\noindent
{\Large\bf Acknowledgments}

\vspace{1cm}

The authors are very grateful to professors I.V.Tuitin, S.V.Ketov, 
our colleagues from the Steklov Mathematical Institute
for the valuable discussions and professors A.A.Slavnov, V.A.Rubakov,
D.Bellisai and M.Matone for the attention to the work. We especially
like to thank V.V.Asadov for the financial support.


\vspace{1cm}

\noindent
{\Large\bf Appendix}

\appendix

\section{N=1 supersymmetric Yang-Mills theories}
\label{susy}
\hspace{\parindent}

The massless N=1 supersymmetric Yang-Mills theory with $SU(N_c)$ gauge group
and $N_f$ matter multiplets is described by the action

\begin{equation}\label{N1_action}
S=\frac{1}{16\pi} \mbox{tr\ Im}\left(\tau \int d^4x d^2\theta\ W^2\right)
+\frac{1}{4} \int d^4x d^4\theta \sum\limits_{A=1}^{N_f}
\left(\phi^{+}_A e^{-2V}\phi^A + \tilde\phi^{+A} e^{2V} \tilde \phi_A
\right)
\end{equation}

\noindent
where the matter superfields $\phi$ and $\tilde\phi$ belong to fundamental
and antifundamental representations of the gauge group $SU(N_c)$.

Here we use the following notations

\begin{eqnarray}
&&A_\mu=e A^a_\mu T^a\qquad \mbox{and so on,}\qquad
\mbox{tr} T^a T^b = \delta^{ab};\nonumber\\
&&\tau = \frac{\theta}{2\pi}+\frac{4\pi i}{e^2};
\end{eqnarray}

\begin{eqnarray}\label{notat}
&&V(x,\theta)=-\frac{i}{2} \bar\theta \gamma^\mu\gamma_5 \theta
A_\mu(x) + i \sqrt{2}(\bar\theta \theta)(\bar\theta\gamma_5\lambda(x))
+ \frac{i}{4} (\bar\theta \theta)^2 D;\nonumber\\
&&W(y,\theta)=\frac{1}{2}(1+\gamma_5)\Big(
i\sqrt{2}\lambda(y) + i\theta D(y)
+\frac{1}{2}\Sigma_{\mu\nu}\theta F_{\mu\nu}(y)\nonumber\\
&&\qquad\qquad\qquad\qquad\qquad\qquad\qquad\qquad\qquad
+\frac{1}{\sqrt{2}}\bar\theta(1+\gamma_5)\theta
\gamma^\mu D_\mu \lambda(y)\Big);\nonumber\\
&&\phi(y,\theta)=\varphi(y)+\sqrt{2}\bar\theta(1+\gamma_5)\psi(y)
+\frac{1}{2}\bar\theta_1(1+\gamma_5)\theta f(y);\nonumber\\
&&y^\mu = x^\mu + \frac{i}{2} \bar\theta \gamma^\mu\gamma_5\theta.
\end{eqnarray}

\noindent
Eliminating auxiliary fields we find that in components the action
(\ref{N1_action}) is written as

\begin{eqnarray}
&&S=\frac{1}{e^2} \mbox{Re\ tr} \int d^4x \left(-\frac{1}{4} F_{\mu\nu}
F^{\mu\nu} - i \bar\lambda(1+\gamma_5)\gamma^\mu D_\mu\lambda
+\frac{\theta e^2}{32\pi^2} F_{\mu\nu}\tilde F_{\mu\nu}
\right)
\nonumber\\
&&+\sum\limits_A\int d^4x \left\{
\vphantom{\frac{1}{2}}
D_\mu \varphi^{+}_A D^\mu \varphi^A
+ D_\mu \tilde\varphi^{+}{}^A D^\mu \tilde\varphi_A
+i \bar\Psi \gamma^\mu D_\mu \Psi
\right.
\nonumber\\
&&
\vphantom{\int}
-i \bar\Psi_A (1-\gamma_5) \lambda\varphi^A
+i \varphi^{+}_A \bar\lambda (1-\gamma_5) \Psi^A
-i \tilde\varphi^{+A}\bar\Psi_A (1+\gamma_5) \lambda
+i \bar\lambda (1-\gamma_5) \Psi^A \tilde\varphi_A
\nonumber\\
&&
\left.
+\frac{1}{2} \left(\varphi^{+}_A T^a \varphi^A
- \tilde\varphi^{+A} T^a \tilde\varphi_A \right)^2
\right\}
\end{eqnarray}

\noindent
where we introduced the Dirac spinor

\begin{equation}
\Psi\equiv \frac{1}{2}
\Big[(1+\gamma_5)\psi+(1-\gamma_5)\tilde\psi\Big]
\end{equation}

In the massless case the action is invariant under the transformations

\begin{eqnarray}
&&U(1)_1:\quad
W(\theta) \to e^{i\alpha} W(e^{-i\alpha\gamma_5}\theta),\quad
\phi(\theta)\to \phi(e^{-i\alpha\gamma_5}\theta),\quad
\tilde\phi(\theta)\to \tilde\phi(e^{-i\alpha\gamma_5}\theta);\nonumber\\
&&U(1)_2:\quad
W(\theta) \to W(\theta),\quad
\phi(\theta)\to e^{i\beta}\phi(\theta),\quad
\tilde\phi(\theta)\to e^{i\beta}\tilde\phi(\theta).
\end{eqnarray}

\noindent
that in components are written as

\begin{eqnarray}\label{symmetry}
U(1)_1: && A_\mu \to A_\mu;\qquad \varphi \to \varphi;\qquad
\tilde\varphi\to\tilde\varphi;\nonumber\\
&&\lambda\to e^{i\alpha\gamma_5}\lambda;\qquad
\Psi\to e^{-i\alpha\gamma_5}\Psi.\nonumber\\
\nonumber\\
U(1)_2: && A_\mu \to A_\mu;\qquad \varphi \to e^{i\beta}\varphi;\qquad
\tilde\varphi\to e^{i\beta}\tilde\varphi;\nonumber\\
&&\lambda\to\lambda;\qquad
\Psi\to e^{i\beta\gamma_5}\Psi.
\end{eqnarray}

The conservation of corresponding currents

\begin{eqnarray}
&& J^\mu_1 = \bar\lambda^a (1 +\gamma_5) \gamma^\mu \lambda^a
+\sum\limits_A \bar\Psi_A \gamma^\mu \gamma_5 \Psi_A; \nonumber\\
&& J^\mu_2 = -\sum\limits_A \bar\Psi_A \gamma^\mu \gamma_5 \Psi_A
-i \sum\limits_A \Big(\varphi_A^{*}
D^\mu \varphi_A - D^\mu\varphi_A^{*} \varphi_A
+ \tilde\varphi_A^{*}
D^\mu \tilde \varphi_A - D^\mu\tilde \varphi_A^{*} \tilde\varphi_A \Big).
\end{eqnarray}

\noindent
is destroyed at the quantum level by anomalies. In the perturbation theory

\begin{eqnarray}\label{curr}
&&\partial_\mu J^\mu_1 = (-N_f + N_c) \frac{1}{16\pi^2}
\varepsilon^{\mu\nu\alpha\beta} \mbox{tr} F_{\mu\nu} F_{\alpha\beta}
=(N_f-N_c)\frac{1}{16\pi^2} \mbox{Im\ tr}\int d^2\theta\ W^2;
\nonumber\\
&&\partial_\mu J^\mu_2 = N_f \frac{1}{16\pi^2}
\varepsilon^{\mu\nu\alpha\beta} \mbox{tr} F_{\mu\nu} F_{\alpha\beta}
=- N_f\frac{1}{16\pi^2} \mbox{Im\ tr}\int d^2\theta\ W^2.
\end{eqnarray}

Nevertheless, it is possible to construct an anomaly free symmetry. Really,
from (\ref{curr}) we conclude, that

\begin{eqnarray}
&&J^\mu_R \equiv J_1^\mu + \frac{N_f-N_c}{N_f} J_2^\mu
= \bar\lambda^a (1+\gamma_5) \gamma^\mu \lambda^a
+ \frac{N_c}{N_f} \sum\limits_A \bar\Psi_A \gamma^\mu \gamma_5 \Psi_A
\nonumber\\
&&-i \sum\limits_A \left(1-\frac{N_c}{N_f}\right) \Big(\varphi_A^{*}
D^\mu \varphi_A - D^\mu\varphi_A^{*} \varphi_A
+ \tilde\varphi_A^{*}
D^\mu \tilde \varphi_A - D^\mu\tilde \varphi_A^{*} \tilde\varphi_A \Big)
\end{eqnarray}

\noindent
is conserved even at the quantum level.

This current is produced by the transformations

\begin{eqnarray}
U(1)_R: &&
W(\theta) \to e^{i\alpha_R} W \Big(e^{-i\alpha_R\gamma_5}\theta\Big);
\nonumber\\
&& \phi(\theta) \to \mbox{exp}\left(i\alpha_R\frac{N_f-N_c}{N_f}\right)\phi
\Big(e^{-i\alpha_R\gamma_5}\theta\Big);
\nonumber\\
&& \tilde \phi(\theta) \to \mbox{exp}
\left(i\alpha_R\frac{N_f-N_c}{N_f} \right)
\tilde\phi \Big(e^{-i\alpha_R\gamma_5}\theta\Big).
\end{eqnarray}

Below we will also use the combination of $U(1)_1$ and $U(1)_2$ with
$\beta=x\alpha$ in (\ref{symmetry}), i.e.

\begin{eqnarray}
U(1)_x: &&W(\theta) \to e^{i\alpha} W(e^{-i\alpha\gamma_5}\theta);\nonumber\\
&&\phi(\theta)\to e^{ix\alpha} \phi(e^{-i\alpha\gamma_5}\theta);\nonumber\\
&&\tilde\phi(\theta)\to e^{ix\alpha} \tilde\phi(e^{-i\alpha\gamma_5}\theta).
\end{eqnarray}

\noindent
where $x$ is an arbitrary constant.

In particular, for $x=(N_f-N_c)/N_f$ we obtain $U(1)_R$ transformations;
for $x=0$ - $U(1)_1$ and for $x \to \infty$ (after redefinition
$\alpha \to \alpha/x$) $U(1)_2$.

The corresponding current is

\begin{eqnarray}
&&J^\mu_x \equiv J_1^\mu + x J_2^\mu
= \bar\lambda^a (1+\gamma_5) \gamma^\mu \lambda^a
+ (1-x) \sum\limits_A \bar\Psi_A \gamma^\mu \gamma_5 \Psi_A
\nonumber\\
&&-i \sum\limits_A x \Big(\varphi_A^{*}
D^\mu \varphi_A - D^\mu\varphi_A^{*} \varphi_A
+ \tilde\varphi_A^{*}
D^\mu \tilde \varphi_A - D^\mu\tilde \varphi_A^{*} \tilde\varphi_A \Big).
\end{eqnarray}

\noindent
In the perturbation theory

\begin{eqnarray}
&&\partial_\mu J^\mu_x = \Big(-N_f + N_c + x N_f\Big) \frac{1}{16\pi^2}
\varepsilon^{\mu\nu\alpha\beta} \mbox{tr} F_{\mu\nu} F_{\alpha\beta}
\nonumber\\
&&\qquad\qquad\qquad\qquad\qquad\qquad
= \Big(N_f - N_c - x N_f\Big) \frac{1}{16\pi^2}
\mbox{Im\ tr}\int d^2\theta\ W^2 .
\end{eqnarray}


\section{Structure of instanton corrections versus collective coordinate
measure transformation law}\label{instanton}
\hspace{\parindent}

In order to define a structure of effective potential we will
calculate anomalies by 2 different ways and compare
the results. \footnote{This approach was first used in \cite{tmf} for N=2
supersymmetric SU(2) Yang-Mills theory.} The action is invariant under
$U(1)_1\times U(1)_2$ group. However, it is more convenient to investigate
the anomaly of $U(1)_x$ symmetry, constructed in Appendix \ref{susy}.

Performing $U(1)_x$ transformation in the effective action we obtain

\begin{equation}\label{N1anomaly1}
\langle\partial_\mu J^\mu_x\rangle =-\frac{\partial\Gamma}{\partial\alpha}
=- \mbox{Im} \int d^2\theta \Big(2w-2\frac{\partial w}{\partial S} S
- x \frac{\partial w}{\partial v} v\Big),
\end{equation}

\noindent
where we substituted $\phi$ and $\tilde\phi$ by their vacuum expectation
values $v$. (For simplicity we assume, that all $v_i$ are equal; a brief
review of the moduli space structure is given in the Appendix \ref{moduli}.)

On the other hand, the anomaly can be found from the transformation law of
the collective coordinate measure.

At the one-instanton level in this case there are $4+N_c^2$ bose zero modes,
$2 N_c$ gluino zero modes (corresponding to supersymmetric ($\epsilon_a$)
and superconformal ($\beta_a$) transformations) and $2N_f$ zero modes for
matter multiplets (supersymmetry $\epsilon_A$). Each zero mode should be
removed by integration over the corresponding collective coordinate. The
measure is written as \cite{cordes}

\begin{eqnarray}\label{measure}
&&d\mu=\nonumber\\
&&=\mbox{const}
\int d^4a \frac{d\rho}{\rho^5} (M\rho)^{4 N_c} d(\mbox{gauge})
\frac{1}{M^{N_c+N_f}\rho^{N_c}}
\prod\limits_{a=1}^{N_c} d\epsilon_a d\bar\beta_a
\prod\limits_{A=1}^{N_f} \frac{d\epsilon_A d\tilde\epsilon_A}{\rho^2 v^2}
\mbox{exp}\left(-\frac{8\pi^2}{e^2} \right) \nonumber\\
&&=\mbox{const} \Lambda^{3N_c-N_f}
\int d^4a d\rho \rho^{3 N_c- 2 N_f-5} \frac{1}{v^{2 N_f}} d(\mbox{gauge})
\prod\limits_{a=1}^{N_c} d\epsilon_a d\bar\beta_a
\prod\limits_{A=1}^{N_f} d\epsilon_A d\tilde\epsilon_A,
\end{eqnarray}

\noindent
where we take into account normalization of all zero modes. The gauge part
and constant factors are written only schematically, because they are not
important in our discussion. As above we need not know the explicit form of
the action in the constant field limit. We should only emphasize, that it
is a dimensionless function of collective coordinates, $\phi$ and, in
principle, $W$. Of course, it is not invariant under $U(1)_x$-transformations

\begin{eqnarray}
&&W  \to e^{i\alpha\gamma_5}W;\qquad\quad\ \
\theta \to e^{-i\alpha\gamma_5}\theta
\nonumber\\
&&\phi \to e^{ix\alpha}\phi;\qquad\qquad\quad
\tilde\phi \to e^{ix\alpha}\tilde\phi;\nonumber\\
&&\epsilon_a \to e^{i\alpha\gamma_5}\epsilon_a;\qquad\qquad\
\beta_a \to e^{i\alpha\gamma_5}\beta_a;\nonumber\\
&&\epsilon_A \to e^{i(x-1)\alpha\gamma_5}\epsilon_A;\qquad
\tilde\epsilon_A \to e^{i(x-1)\alpha\gamma_5}\tilde\epsilon_A;
\nonumber\\
&&\rho \to \rho;\qquad\qquad\qquad\ \ \ a^\mu{} \to a^\mu
\end{eqnarray}

\noindent
as above.

Let us perform an additional substitution

\begin{eqnarray}\label{compens}
&&\theta \to e^{-i\alpha\gamma_5}\theta;\qquad\qquad\quad\ \
x^\mu \to e^{-2i\alpha} x^\mu;\qquad\nonumber\\
&&\epsilon_A \to e^{-ix\alpha\gamma_5}\epsilon_A;\qquad\qquad\
\tilde\epsilon_A \to e^{-ix\alpha\gamma_5}\tilde\epsilon_A;
\qquad\nonumber\\
&&\rho \to e^{-2i\alpha} \rho;\qquad\qquad\qquad
a^\mu \to e^{-2i\alpha} a^\mu,\qquad\nonumber\\
\end{eqnarray}

\noindent
so that the final transformations

\begin{eqnarray}\label{overall}
&&(1+\gamma_5)\epsilon_a \to e^{i\alpha}(1+\gamma_5)\epsilon_a;
\qquad\qquad\ \
(1-\gamma_5)\beta_a \to e^{-i\alpha}(1-\gamma_5)\beta_a;\qquad\nonumber\\
&&(1+\gamma_5)\epsilon_A \to e^{-i\alpha}(1+\gamma_5)\epsilon_A;\qquad\qquad
(1+\gamma_5)\tilde\epsilon_A \to e^{-i\alpha}(1+\gamma_5)
\tilde\epsilon_A;\qquad\nonumber\\
&&\rho \to e^{-2i\alpha} \rho;\qquad\qquad\qquad
a^\mu \to e^{-2i\alpha} a^\mu;\qquad\qquad\qquad
\theta \to e^{-2i\alpha\gamma_5}\theta\qquad
\end{eqnarray}

\noindent
(except for $\theta$) correspond to dimension of the fields. The
dimensionless action would have been invariant, if we had made additional
rotation

\begin{equation}\label{lack}
v \to e^{i(2-x)\alpha} v;\qquad
W \to e^{2i\alpha\gamma_5} W.
\end{equation}

\noindent
However, we can not make it because $v$ and $W$ are not collective
coordinates (and, therefore, integration variables). It means, that
under (\ref{overall})

\begin{eqnarray}
&&S(v,W) \to S(e^{i(x-2)\alpha} v, e^{-2i\alpha\gamma_5} W);\nonumber\\
&&d\mu(v) \to \mbox{exp}\left[i\alpha\Big(-2(3N_c-2N_f) -2N_fx + 2N_f(x-2)
- 2N_f\Big)\right] d\mu(e^{i(x-2)\alpha}v)  \nonumber\\
&&=\mbox{exp}\left[i\alpha\Big(-2(3N_c-N_f)\Big)\right]
d\mu(e^{i(x-2)\alpha}v).
\end{eqnarray}

\noindent
It is quite evident, that the $n$-instanton collective coordinate
measure is transformed as

\begin{equation}\label{measurecontribution}
d\mu(v) \to \mbox{exp}\left[i\alpha\Big(-2n(3N_c-N_f) \Big)
\right] d\mu(e^{i(x-2)\alpha}v).
\end{equation}

\noindent
Moreover, we should also perform the inverse substitution in the remaining
integral (see the definition of the superpotential)

\begin{equation}\label{lastint}
\int d^4x d^2\theta \to e^{4i\alpha}\int d^4x d^2\theta,
\end{equation}

\noindent
so that finally from (\ref{lack}), (\ref{measurecontribution}) and
(\ref{lastint}) we conclude, that

\begin{equation}
w(v,W) \to \mbox{exp}\left[i\alpha\Big(-2n(3N_c-N_f)+4 \Big)\right]
w(e^{i(x-2)\alpha} v, e^{-2i\alpha\gamma_5} W).
\end{equation}

\noindent
Taking into account that the action contains only $(1+\gamma_5) W$, we find
the anomaly to be

\begin{eqnarray}\label{N1anomaly2}
&&\langle\partial_\mu J^\mu_R\rangle
=-\left. \frac{\partial\Gamma}{\partial\alpha}\right|_{\alpha=0}=
\nonumber\\
&&\qquad\qquad =\mbox{Im} \int d^2\theta \left(- 2n(3N_c-N_f)
+ (-2+x) v\frac{\partial}{\partial v} - 2 W\frac{\partial}{\partial W}
+ 4 \right) w. \qquad
\end{eqnarray}

\noindent
Comparing (\ref{N1anomaly1}) and (\ref{N1anomaly2}), we obtain the
following equation for $n$-instanton contribution to the superpotential:

\begin{equation}
\Big(2v\frac{\partial}{\partial v}
+ 3 W \frac{\partial}{\partial W} - 6 \Big) w^{(n)}
= - 2 n (3N_c-N_f) w^{(n)}.
\end{equation}

\noindent
It is easily verified, that the solution is

\begin{equation}\label{superpotential0}
w^{(n)}
=  W^2 g_n\left(\frac{v^3}{W^2}\right)
\left(\frac{\Lambda}{v}\right)^{n(3N_c-N_f)}
=  S g_n\left(\frac{v^3}{S}\right)
\left(\frac{\Lambda}{v}\right)^{n(3N_c-N_f)}
\end{equation}

\noindent
where $g_n$ is an arbitrary function. Its explicit form can be found
from the relation between perturbative and exact anomalies.

Of course, the result (\ref{superpotential0}) is in a complete agreement
with dimensional arguments and does not depend on the particular choice of
symmetry (i.e. $x$).


\section{The classical moduli spaces of N=1 supersymmetric theories}
\label{moduli}
\hspace{\parindent}

To describe the vacuum states it is convenient to introduce two
$N_f\times N_c$ matrixes of the form

\begin{equation}
\phi \equiv \left(\phi^1, \phi^2,\ldots,\phi^{N_f}\right);\qquad
\tilde \phi \equiv \left(\tilde \phi_1, \tilde \phi_2,\ldots,
\tilde \phi_{N_f}\right).
\end{equation}

\noindent
(Their rows correspond to different values of color index.) The energy is
minimal if $\phi=\tilde \phi\equiv v$. Performing rotations in the color
and flavor spaces we can always reduce the matrix $v$ to the form

\begin{equation}
v=\left(
\begin{array}{cccc}
v_1    & 0      & \ldots & 0\\
0      & v_2    & \ldots & 0\\
\ldots & \ldots & \ldots & \ldots\\
0      & 0      & \ldots & v_{N_f}\\
0      & 0      & \ldots & 0\\
\ldots & \ldots & \ldots & \ldots
\end{array}
\right)
\end{equation}

\noindent
if $N_f < N_c$ and

\begin{equation}
v=\left(
\begin{array}{ccccccc}
v_1    & 0      & \ldots & 0       & 0      & \ldots \\
0      & v_2    & \ldots & 0       & 0      & \ldots \\
\ldots & \ldots & \ldots & \ldots  & \ldots & \ldots \\
0      & 0      & \ldots & v_{N_c} & 0      & \ldots
\end{array}
\right)
\end{equation}

\noindent
if $N_f > N_c$.

\vspace{5mm}

\noindent
1. $N_f < N_c$.

In the generic point the gauge group $SU(N_c)$ is broken down to
$SU(N_f-N_c)$. Therefore,

\begin{equation}
\Big(N_c^2-1\Big)-\Big((N_c-N_f)^2-1\Big) = 2 N_c N_f - N_f^2
\end{equation}

\noindent
chiral superfields are eaten up by super-Higgs mechanism. Taking into account
that originally there are $2 N_c N_f$ chiral matter superfields, we conclude
that only

\begin{equation}
2N_C N_f-\Big(2 N_c N_f - N_f^2\Big)=N_f^2
\end{equation}

\noindent
ones remain massless.

The flat direction can be described in the gauge invariant way by
$N_f^2$ composite chiral superfields

\begin{equation}
M_A{}^B = \tilde \phi_{Aa} \phi^{Ba}.
\end{equation}

\noindent
(Here $a$ denotes a color index.)

\vspace{5mm}

\noindent
2. $ N_f \ge N_c$.

If the number of flavors is equal to or larger than the number of colors,
the original gauge group is completely broken in the generic point.
Therefore, the number of remaining massless chiral superfields is

\begin{equation}
2 N_c N_f -\Big(N_c^2-1 \Big) = 2 N_c N_f - N_c^2 +1.
\end{equation}

In this case the gauge invariant description is provided by "mesons"

\begin{equation}
M_A{}^B = \tilde \phi_{Aa} \phi^{Ba}
\end{equation}

\noindent
and "barions"

\begin{eqnarray}
&&B_{A_{N_c+1}A_{N_c+2}\ldots A_{N_f}} = \frac{1}{N_c!}
\varepsilon_{A_1 A_2\ldots A_{N_f}} \varepsilon^{a_1 a_2\ldots a_{N_c}}
\phi^{A_1 a_1} \phi^{A_2 a_2} \ldots \phi^{A_{N_c} a_{N_c}};\nonumber\\
&&\tilde B^{A_{N_c+1}A_{N_c+2}\ldots A_{N_f}} = \frac{1}{N_c!}
\varepsilon^{A_1 A_2\ldots A_{N_f}} \varepsilon^{a_1 a_2\ldots a_{N_c}}
\phi_{A_1 a_1} \phi_{A_2 a_2} \ldots \phi_{A_{N_c} a_{N_c}}.
\end{eqnarray}

\noindent
However, their overall number is greater than $2 N_c N_f - N_c^2 +1$.
The matter is that at the classical level these fields are not
independent and satisfy some constraints. For example, if $N_f = N_c$
the number of massless superfields is $N_f^2+1$ while $N_M+N_B=N_f^2+2$.
The constraint eliminating the redundant chiral variable is

\begin{equation}
\mbox{det} M = \tilde B B.
\end{equation}

Similarly, for $N_f=N_c+1$

\begin{eqnarray}
&& B_A M_A{}^B = M_B{}^A \tilde B_A =0;\nonumber\\
&&\mbox{det} M \Big(M^{-1}\Big)_A{}^B = B_A \tilde B^B.
\end{eqnarray}

However, at the quantum level these constraints are violated by
instanton corrections and are no longer valid \cite{seiberg2}.


\section{The gauge invariant form of parameter z}\label{derive_z}
\hspace{\parindent}

If $N_f < N_c$, the only gauge invariant parameter of $v^{2N_f}$ order
is det $M$, so that

\begin{equation}
z=\frac{\Lambda^{3N_c-N_f}}{\mbox{det} M\ S^{N_c-N_f}}.
\end{equation}

For $N_f \ge N_c$ the moduli space is parametrized by mesons $M_A{}^B$ and
barions $B_{B_1\ldots B_{N_f-N_c}}$, $\tilde B^{A_1\ldots A_{N_f-N_c}}$,
satisfying some classical constrains. At the quantum level these constrains
are broken by instanton corrections. In the effective action approach the
modifications should be produced automatically. It can be achieved by
integrating out the $S$-superfield. (In particular, for $N_f=N_c$ $S$
is a natural Lagrange multiplier). The result should have the following
form \cite{peskin}:

\begin{eqnarray}\label{effective}
&& \mbox{det} M - \tilde B B = \mbox{const}\ \Lambda^{2 N_f},\qquad N_f=N_c;
\nonumber\\
\nonumber\\
&& w_{eff} = \mbox{const}\
\Lambda^{\textstyle - \frac{3N_c-N_f}{N_f-N_c}}
\Big(\mbox{det} M - (\tilde B^{A_1 A_2\ldots A_{N_f-N_c}} M_{A_1}{}^{B_1}
M_{A_2}{}^{B_2} \ldots M_{A_{N_f-N_c}}{}^{B_{N_f-N_c}}\nonumber\\
&&\times B_{B_1 B_2\ldots B_{N_f-N_c}})
\Big)^{\textstyle - \frac{1}{N_f-N_c}}+h.c.,
\qquad N_f > N_c.
\end{eqnarray}

\noindent
It can be achieved if and only if $v^{2N_f}$ is substituted by

\begin{equation}
\mbox{det} M - (\tilde B^{A_1 A_2\ldots A_{N_f-N_c}} M_{A_1}{}^{B_1}
M_{A_2}{}^{B_2} \ldots M_{A_{N_f-N_c}}{}^{B_{N_f-N_c}}
B_{B_1 B_2\ldots B_{N_f-N_c}}),
\end{equation}

\noindent
so that finally

\begin{equation}
z=\frac{\Lambda^{3N_c-N_f}S^{N_f-N_c}}{
\mbox{det} M - (\tilde B^{A_1 A_2\ldots A_{N_f-N_c}} M_{A_1}{}^{B_1}
M_{A_2}{}^{B_2} \ldots M_{A_{N_f-N_c}}{}^{B_{N_f-N_c}}
B_{B_1 B_2\ldots B_{N_f-N_c}})}.
\end{equation}

We would like to mention, that in the presented approach (\ref{effective})
certainly contains multiinstanton corrections, that contribute to the overall
constant factor in the RHS.


\end{document}